# Superconducting State and Phase Transitions

Yury Mnyukh [1], Vitaly J. Vodyanoy [2,*]

[1]Chemistry Department and Radiation and Solid State Laboratory, New York University, New York, NY USA
[2]Biosensors Laboratory, Alabama Micro/Nano Science and Technology Center, Auburn University, Auburn Alabama USA

**Abstract** From the days when superconductivity was discovered its science was entangled by the unresolved problem of the relationship between superconducting state, its crystal structure and its phase transitions. The problem was exacerbated by the adjacent scientific area – solid state phase transitions – that offered to divide phase transitions into first and second order. Adding to that the conclusion (turned out erroneous) on the structural identity of the superconducting and normal phases, the whole issue was uncertain and confusing. This article straights it out. The first step in that direction was to show that the phase transitions attested previously as a second order are first order. It is done by proving that *all* solid-state phase transitions are first order. This was achieved by revealing that (a) introduction of the second-order type by Ehrenfest was baseless on the ground that the "heat capacity λ-anomaly" in liquid He was actually latent heat, (b) the Justy and Laue thermodynamic consideration that the second-order type cannot exist was rigorous and had not to be neglected, and (c) not a single well-proven second-order phase transition in solid state was ever found. Thermodynamic arguments are presented that being a first-order phase transition means crystal rearrangement, and that the universal *nucleation-and-growth* mechanism of phase transitions is the way it to materialize. In particular case of normal – superconducting phase transitions we have shown that (1) the normal and superconducting crystal structures are not identical, (2) the phase transition between them proceeds by a structural rearrangement, (3) the *epitaxial* version of the *nucleation-and-growth* is its molecular mechanism, (4) because the mechanism is universal toward all kinds of solid-state phase transitions, there was no basis to assume that it leads to description of superconductivity. Elimination of the identified wrong concepts and misinterpretations will accelerate the advancement of the science of superconductivity toward creation of efficient inexpensive high-temperature superconductors.

## 1. Introduction

Investigations of the crystal structure and phase transitions in the context of superconductivity basically follow two different lines. More frequently it is an analysis of the structural type, structural features and polymorphic forms of the known or potential superconductors, which they exhibit under different temperatures, pressures, substitutions or dopings. This line is aimed at empirical search for better superconductors. The second line is investigations of the normal – superconducting phase transitions at the temperature where they occur. The present article is of the latter type.

Immediately after the discovery of superconductivity in 1911 by Kamerlingh-Onnes the question on *how* superconducting state emerges from the higher-temperature "normal" non-superconducting state arose. Finding the answer was perceived to be a major step toward understanding the superconducting state itself. With that in mind, it had to be first established whether superconducting phase has its individual crystal structure. A science historian P. F. Dahl described the relevant state of affairs of those days in his 1992 book [1]: "Onnes then returned to the crucial question raised first by Langevin at Brussels in 1911 regarding evidence for a change of phase at the transition point. An experimental search for a latent heat of transformation was planned by Onnes with the assistance of the American Leo Danna … in 1922-1923. … In any case, Onnes seems by now to have been more inclined to view the transition in terms of Bridgman's polymorphic change. This possibility was, however, negated by a decisive experiment concluded by Keesom shortly before the 1924

Solvay Conference. Analysis of Debye-Scherrer X-ray diffraction patterns revealed no change in the crystal lattice of lead ...".

It will be shown in the present article that the conclusion about crystallographic identity of pre-superconducting and superconducting phases was in error. It had and is still has a lasting detrimental effect on the investigation of superconductivity. Its general acceptance relied on an experiment performed on only one superconductor without later verification with improving techniques and without validation on other superconductors. It served as a basis to erroneously categorize these transitions as "second order" after Ehrenfest [2] introduced his first/second order classification in 1933.

All superconductors known at that time are now recognized by most to exhibit first order phase transition, but the newer ones are sorted out in both ways, and even hovering somewhere in between. Assigning the "order" is being done in a formal manner, using unreliable criteria, and without specifying the physical content behind each "order". Missing is a familiarity with the solid-state phase transitions in general, an adjacent area of solid-state physics on its own rights. Our purpose is to present the physical picture of a *normal – superconducting* phase transition and its relation to the superconducting state in terms of the general *nucleation-and-growth* molecular mechanism of structural rearrangements. This mechanism is described in Section 5 in sufficient detail. The "order" issue needs to be analyzed first.

## 2. First-Order and Second-Order Phase Transitions

What are a "phase" and a "phase transition"? "A *phase, in the solid state, is characterized by its structure. A solid-state phase transition is therefore a transition involving a change of structure, which can be specified in geometrical terms*" [3]. There are only two conceivable ways the phase transition may occur without being at variance with thermodynamics. An infinitesimal change of a controlling parameter (dT in case of temperature) may produce either (*A*) emerging of an infinitesimal *quantity* of the new phase with its structure and properties changed by finite values, or (*B*) a cooperative infinitesimal "*qualitative*" physical change throughout the whole macroscopic bulk [4]. It is imperative to realize that not any third way can exist. Thus, finite changes by "distortion" or "deformation" of the crystal, or "displacement" of its particles are not possible.

There is no guarantee that both versions *'A'* and *'B'* materialize in nature. However, in 1933 Ehrenfest formally classified phase transitions by *first-order* and *second-order* (see Section 3). The validity of the classification was disputed by Justi and Laue by stating that there is no thermodynamic or experimental justification for second-order phase transitions. Their objections were ignored and forgotten over the ensuing decades (see Section 6).

In 1935-1937 Landau [5, 6] developed the theory of *second-order* phase transitions. But he acknowledged that transitions between different crystal modifications are "usually" *first-order*, when "jump-like rearrangement takes place and state of the matter changes abruptly"; at that, a latent heat is absorbed or released, symmetries of the phases are not related and overheating or overcooling is "possible". As for *second-order* phase transitions, they "may also exist", but no incontrovertible evidence of their existence was presented. (It should be noted that the expression "may exist" implicitly allows something not to exist either). *Second-order* phase transitions must be cooperative and occur without overheating or overcooling at fixed "critical points" where only the crystal symmetry changes, but structural change is infinitesimal. It is specifically emphasized that "second order phase transitions, as distinct from first-order transitions, are not accompanied by release or absorption of heat". These features were in accord with those by Ehrenfest, namely, no latent heat, no entropy change, no volume change, and no phase coexistence.

Since then it has become universally accepted that there are "jump-like discontinuous" *first-order* phase transitions, as well as "continuous" *second-order* phase transitions without "jumps". The latter fit well the thermodynamic requirement *'B'*, leaving *first-order* phase transitions to be associated with the requirement *'A'*. The theoretical physicists were so preoccupied with second-order phase transitions, even trying to treat first-order transitions as second order that neglected to take a close look at the first-order transitions. If they did, the need to comply with the requirement *'A'* would be noticed. That requirement contains essential information on their molecular mechanism. In a sense, they are also "continuous". Contrary to the everybody's conviction, instant jump-like macroscopic changes do not occur. The transitions proceed by successive transfer of *infinitesimal* amounts of the material from initial to the resulting phase, which on the microscopic scale means molecule-by-molecule. During that process the two phases inevitably coexist. It will be shown in Section 4 that the experimentally discovered mechanism exactly fits the requirement *'A'*.

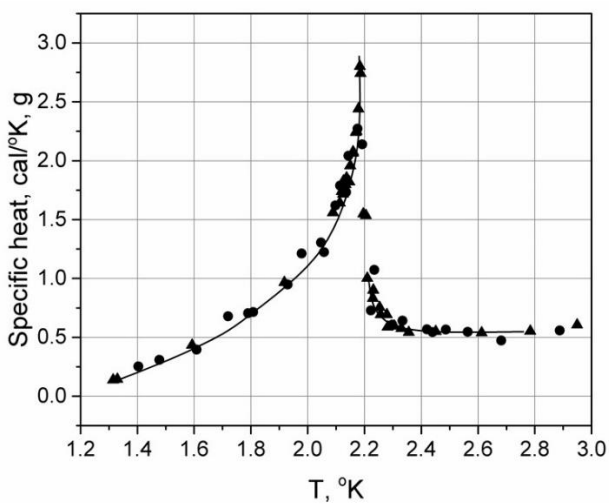

Fig. 1. The "λ-Anomaly" recorded in heat capacity measurements of liquid He phase transition (adopted from [7]).

Without going into the Landau theory itself, there were several shortcomings in its presentation:

(a) He had not answered the arguments of the contemporaries, M. Laue among them, that second-order phase transitions do not - and cannot – exist.

(b) The only examples used in illustration of second-order phase transitions, $NH_4Cl$ and $BaTiO_3$, turned out to be *first order*.

(c) The theory was unable to explain the so called "heat capacity λ-anomalies" which were regarded being second order. The fact that they also appeared in first-order phase transitions was not addressed either.

(d) Overheating and overcooling in first-order phase transitions are not only "possible", but inevitable, considering that they will not proceed when the free energies $F_1$ and $F_2$ are equal and, therefore, the driving force is absent.

(e) First-order phase transitions were described as "abrupt jumps-like rearrangements", apparently assuming them to comprise a finite bulk of matter at a time. The same is seen from allowing first-order phase transitions to sometimes occur without hysteresis, in which cases they would occur at the single temperature point $T_o$ when $F_1 = F_2$, involving instantly the whole crystal – in contradiction with the condition *'A'*.

(f) When claiming that second-order phase transitions "may also exist", no estimates, or even arguments, were presented that they could be more energy advantageous in some cases. However, a strong argument can be raised that they can never materialize, considering that the first-order phase transitions (matching the condition *'A'*) would be always preferable, requiring energy to relocate only one molecule at a time, rather than the myriads of molecules at a time as a cooperative process of second-order phase transitions assumes.

## 3. Ehrenfest Classification: Construction on Quick Sand

### 3.1. "λ-Anomaly" in Liquid Helium and the Idea of Second-Order Phase Transitions

In 1932 Keesom and coworkers [7] discovered the "heat capacity λ-anomaly" in liquid He phase transition (Fig. 1). This apparent new type of phase transition was the reason for Ehrenfest to put forward his classification. Here are a few excerpts from the 1998 article [8] where the events were summarized: "It is important to note a general scheme was generated on the basis of just one "unusual" case, liquid helium. It seemed probable that the scheme would be applicable to other known systems, such as superconductors … The liquid-helium lambda transition became one of the most important cases in the study of critical phenomena – its true (logarithmic) nature was not understood for more than ten years … The Ehrenfest scheme was then extended to include such singularities, most notably by A. Brain Pippard in 1957, with widespread acceptance. During the 1960's these logarithmic infinities were the focus of the investigation of "scaling" by Leo Kadanoff, B. Widom and others. By the 1970s, a radically simplified binary classification of phase transitions into "first-order" and "continuous" transitions was increasingly adopted." To the voluminous theoretical efforts listed in the last excerpt, the 1982 Nobel Prize to Kenneth Wilson for his scaling theory of second-order phase transitions had been added. Unfortunately, all that activity descends from a wrong interpretation of the "λ-anomaly" in liquid He: it is not heat capacity, but the latent heat of the first-order phase transition. Allegorically, the tower was erected on quicksand.

### 3.2. "Heat Capacity λ-Anomalies" in Solid-State Phase Transitions

It is not clear why Ehrenfest did not invoke the "heat capacity λ-anomaly" in NH4Cl phase transition, discovered by F. Simon in 1922 [9], for it exhibited the same peculiar λ-shape as that in liquid He. Later on, that λ-peak was reproduced many times by other workers and numerous similar cases were also reported. Thus, more than 30 experimental λ-peaks presented as "specific heat $C_P$ of [substance] vs. temperature T" were shown in the 1978 book by Parsonage and Staveley [10]. Theorists were unable to account for the phenomenon. P.W. Anderson wrote [11]: "Landau, just before his death, nominated [lambda-anomalies] as the most important as yet unsolved problem in theoretical physics, and many of us agreed with him… Experimental observations of singular behavior at critical points… multiplied as years went on… For instance, it have been observed that magnetization of ferromagnets and antiferromagnets appeared to vanish roughly as $(T_C-T)^{1/3}$ near the Curie point, and that the λ-point had a roughly logarithmitic specific heat $(T-T_C)^0$ nominally". Feynman stated [12] that one of the challenges of theoretical physics today is to find an exact theoretical description of the character of the specific heat near the Curie transition - an intriguing problem which has not yet been solved.

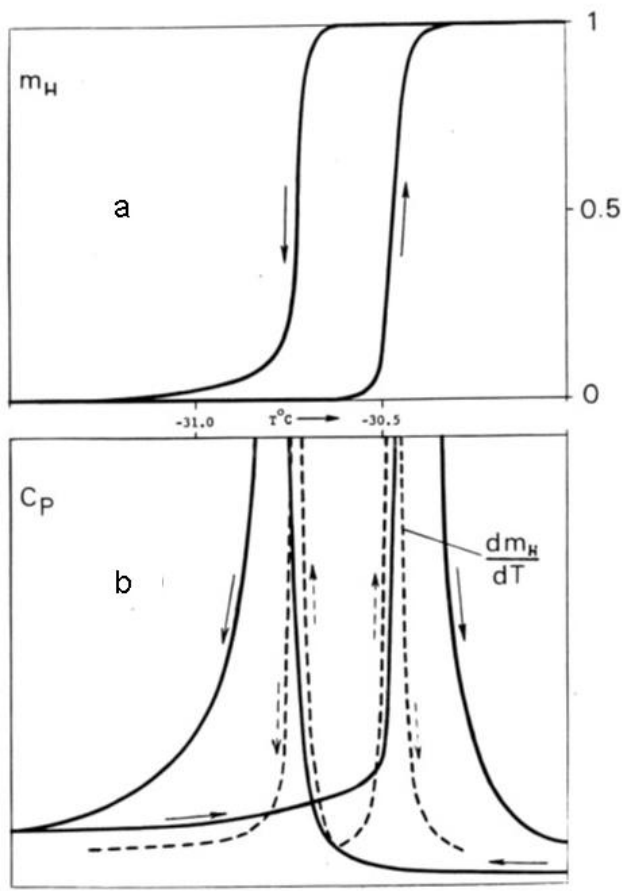

Fig. 2. Phase transition in NH4Cl.. a. The hysteresis loop by Dinichert [22] represents a mass fraction of high-temperature phase, mH, in the two-phase, L+H, range of transition; mL+mH =1. b. Solid lines: The λ-peaks from calorimetric measurements by Extermann and Weigle [21]. The peaks are a subject of hysteresis. The plots 'a' and 'b' are positioned under one another on the same temperature scale to make it evident that the shape of the peaks is proportional to a first derivative of the $m_H(T)$ (dotted ).

The problem was hidden in the erroneous interpretation of available experimental data. There were three main reasons for that theoretical impasse. (1) Attention was not paid to the fact that λ-peaks were actually observed in first-, and not second-order phase transitions. (2) The first-order phase transitions exhibited latent heat, but it was mistaken for heat capacity. (3) An important limitation of the adiabatic calorimetry utilized in the measurements was unnoticed.

### 3.3. The "λ-Anomaly" in NH4Cl is Latent Heat of First-Order Phase Transition

The canonical case of "specific heat λ-anomaly" in NH4Cl around -30.6 °C has been reexamined [13] (also Section 3.4.6 and Appendix 2 in [14]). That case is of a special significance. It was the first where a λ-peak in specific heat measurements through a solid-state phase transition was reported and the only example used by Landau in his original articles on the theory of continuous second-order phase transitions. This particular phase transition was a subject of numerous studies by different experimental techniques and considered most thoroughly investigated. In every calorimetric work (e. g. [15-21] a sharp λ-peak was recorded; neither author expressed doubts in its *heat capacity* nature. The transition has been designated as a *cooperative order-disorder phase transition of the lambda type* and used to exemplify such a type of phase transitions. No one claimed that the λ-anomaly was understood.

It should be noted that many of the mentioned calorimetric studies were undertaken well after 1942 when the experimental work by Dinichert [22] was published. His work revealed that the transition in NH4Cl was spread over a temperature range where only mass fractions $m_L$ and $m_H$ of the two distinct L (low-temperature) and H (high-temperature) coexisting phases were changing, producing "sigmoid"-shaped curves. The direct and reverse runs formed a hysteresis loop shown in Fig. 2**a**. The fact that the phase transition is first-order was incontrovertible, but not identified as such.

In Fig. 2**b** the Dinichert's data were compared with the calorimetric measurements by Extermann and Weigle [17]. The latter exhibited "anomalies of heat capacity" (as the authors called the λ-peaks) and the hysteresis of the λ-peaks. Because of the hysteresis, it had to be evident (but was not) that the λ-peaks cannot be of a heat capacity, considering that heat capacity is a *unique function* of temperature. The graphs 'a' and 'b' are positioned under one another on the same temperature scale to reveal that the shape and location of the peaks are very close to the *first derivative* of the $m_H(T)$ (dotted curves). It remains to note that the *latent heat* of the phase transition must be proportional to $dm_H/dT$. Thus, the latent heat of the first-order phase transition, lost in the numerous calorimetric studies, was found, simultaneously eliminating the great theoretical mystery.

### 3.4. Limitation of Adiabatic Calorimetry

The Dinichert's work had not changed the interpretation of the λ-peak from "heat capacity" to "latent heat", even though all the attributes of a structural first-order phase transition have there been exposed. A partial explanation of that is: one had to go beyond the abstract term "first order" and its superficial indicators "jump/no jump" or "latent heat/no latent heat", straight to its physical essence, which is a *quantitative* transfer of the matter at interfaces between the competing crystal phases. And that was not the case. But there was also second reason, and it was hidden in the calorimetric technique itself.

The goal of numerous calorimetric studies of λ-peaks in $NH_4Cl$ and other substances was to delineate the shape of these peaks with the greatest possible precision. An adiabatic calorimetry, it seemed, suited best to achieve it. The adiabatic calorimeters, however, are only "one-way" instruments in the sense the measurements can be carried out only as a function of increasing temperature. In the case under consideration, however, it was vital to perform both temperature-ascending and descending runs – otherwise the existence of hysteresis would not be detected. For example, in [16] the transition in $NH_4Cl$ was claimed to occur at fixed temperature $T_\lambda = 245.502 \pm 0.004$ K defined as a position of the λ-peak. The high precision of measurements was useless: that $T_\lambda$ exceeded $T_o$ by $3°$.

The results by Extermann and Weigle were not typical. The kind of calorimetry they utilized permitted both ascending and descending runs. That was a significant advantage over the adiabatic calorimetry used by others in the subsequent years. But there was also a shortcoming both in their and the adiabatic calorimetry techniques resulted in the unnoticed error in the presentation of the λ-peaks in Fig. 2**b**: the *exothermic latent heat* peak in the descending run had to be *negative* (looking downward)**.** Adiabatic calorimetry, in spite of all its precision, could not detect that.

### 3.5. Straightforward Test of the Heat Effect in $NH_4Cl$

Differential scanning calorimetry (DSC) is free of the above shortcomings [23]. Carrying out temperature descending runs with DSC is as easy as ascending runs. Most importantly, it displays endothermic and exothermic peaks with *opposite* signs in the chart recordings, which results from the manner the signal is measured. If the λ-peak in $NH_4Cl$ is a *latent heat* of phase transition, as was concluded above, the peak in a descending run must be exothermic and look downward. Our strip-chart recordings made with a Perkin-Elmer DSC-1B instrument immediately revealed [13] that the peak acquires opposite sign in the reverse run (Fig. 3). Its hysteresis was also unveiled.

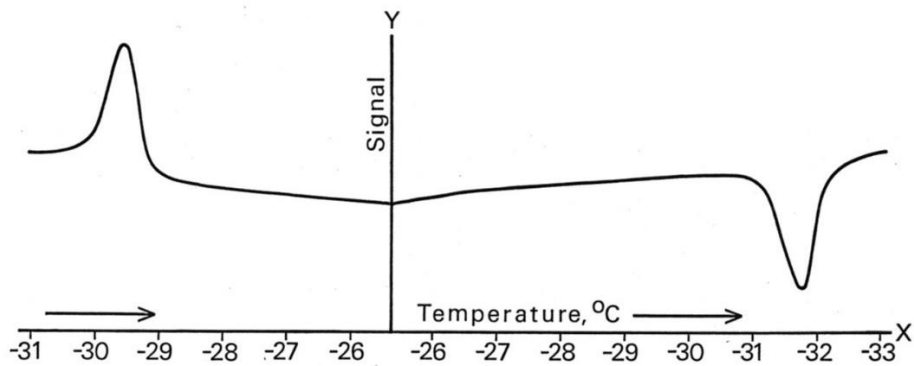

Fig. 3. The actual DSC recording of NH₄Cl phase transition cycle, displaying temperature-ascending and descending peaks as endothermic and exothermic accordingly, thus delivering visual proof of their latent heat nature.

### 3.6. Consequential Mistake: "λ-Anomaly" in Liquid Helium is Latent Heat of Phase Transition, not Heat Capacity

It is not realistic to continue assigning a *heat capacity* to the λ-peak in liquid He, while the same peculiar peak in NH₄Cl is proven to be a *latent heat*. Looking into the experimental techniques utilized in the investigations of the liquid He phase transition we find that, indeed, it is the adiabatic calorimetry with its inherent limitations that was used. The critical claim that latent heat is absent in that transition was made on the basis that it was not observed in passing along the λ-curve. The irony is that the whole λ-peak is none other than a latent heat.

The distinction between phase transitions in solid and liquid states is not so wide as it may seem. In liquid *He* we deal with the *liquid polymorphism* – a phenomenon possibly unknown when its λ-peak was discovered, but is known presently [24]. The liquid is not a completely disordered state. At any given moment it consists of multiple tiny clusters of approximately closely packed particles. A certain kind of a *short-range* order within the clusters is preserved, analogous to the *long-range* order of crystals. Change in the manner the particles are packed in the clusters with temperature (or pressure) is similar to polymorphic phase transitions in crystals. *We maintain that measurements of the heat effect in the liquid He phase transition with an appropriate calorimeter (such as differential scanning calorimeter) will produce the peak in cooling and heating runs as looking in opposite directions* (as in Fig. 3)*, thus exhibiting its latent heat nature. Simultaneously, its hysteresis will be revealed, in which case the λ-peak in the cooling run will be shifted from its current temperature 2.19 K toward 1 K.*

The physical phenomenon used by Ehrenfest to introduce the *second-order* type of phase transitions was not existed. Neither the subsequent voluminous work over many years on shape delineation and analytic description of the "specific heat λ-anomaly" has any scientific value.

### 4. Nucleation-and-Growth Mechanism of Solid-State Phase Transitions

The *nucleation-and-growth* molecular mechanism of solid-state phase transitions was derived by Mnyukh in 1971-1979 from the results of systematic experimental studies [25-38], later summarized in [14] and finalized in [39-45]. Solid-state phase transitions were found to be a *crystal growth,* very similar to the crystal growth from the liquid phase, but involving a distinctive kind of nucleation and a specific *contact* structure of interfaces. It was demonstrated that the *nucleation-and growth* mechanism is general to all kinds of solid-state phase transitions, including ferromagnetic, ferroelectric and order-disorder. It will be now briefly described, for it covers the *normal – superconducting* phase transitions as well.

4.1. **Nucleation in Microcavities. Hysteresis**

Nucleation of the alternative phase in solid-state reactions differs in all respects from the theoretical-born fluctuation-based statistical process described in the Landau and Lifshitz classical textbook ( Section 150 in [6] ). Nucleation in a given crystal is pre-determined as to its location and temperature. It would not occur at all in perfect crystals, but real crystals are never perfect. The nucleation sites are located in specific crystal defects – microcavities of a certain optimum size. These defects contain information on the condition (*e.g.*, temperature) of their activation. Nucleation lags are not the same in different defects, but are finite. *Hysteresis is a necessary component if the phase transition mechanism*. The "ideal phase transition" without hysteresis at the temperature $T_o$ ($F_1 = F_2$) cannot occur, considering that the thermodynamic driving force $\Delta F = f(T-T_o) = 0$ at $T = T_o$. Taking into account that the nucleation lags occur in both directions of the phase transition, some finite *hysteresis* of phase transitions is inevitable. Its value, however, is not exactly reproducible. This is the hysteresis of *initiation* of phase transition in a whole crystal or its macroscopic parts the researchers usually encounter.

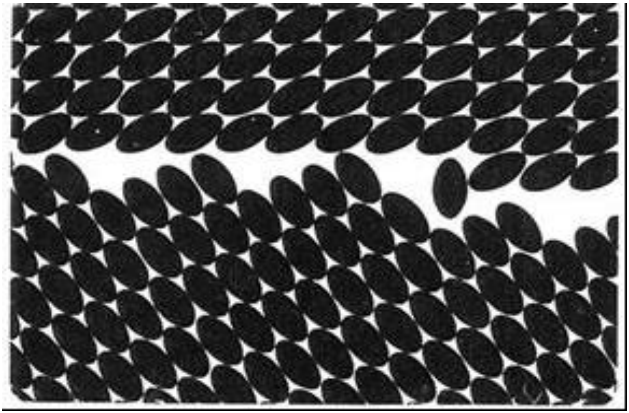

Fig. 4. Model of a contact interface. Phase transition proceeds by molecule-by-molecule building up new layers of the emerging (upper part) phase. It is in accord with the thermodynamic condition 'A' that requires emerging only an "infinitesimal" quantity of the new phase at a time.

Less prominent is another type of nucleation. Proceeding of the phase transition after its initiation in a crystal requires the presence of a different type of microcavities in sufficient concentration. These are too small to initiate the phase transition but bigger than a single vacancy. If their concentration is lower than a certain critical value, the original phase remains practically stable at all temperatures, while being theoretically unstable according to thermodynamics.

4.2. **Structural Rearrangement at Contact Interface. Phase Coexistence**

A solid-state phase transition is an intrinsically *local* process. It proceeds by "molecule-by-molecule" structural rearrangement at interfaces only, while the rest bulk of the original and the emerged phases remain static (Fig.4). This mechanism is in compliance with the requirement *'A'* in Section 2, namely, no macroscopic "jumps" occur during the phase transition. The seeming "jumps" are simply the differences between physical properties of the initial and resultant phases, appearing in the experiments as "jumps" when the transition range is either narrow enough, or passed quickly, or both.

The coexistence of the phases during phase transition is self-evident.

**4.3. Epitaxial Phase Transitions**

Fig. 4 illustrates a case when crystal orientations of the phases are not related. This is because the nuclei that initiate phase transitions in crystal defects have, in principle, arbitrary orientations. However, a structural *orientation relationship* (OR) is frequently observed. That does not mean these transitions occur by a "deformation" or "distortion", involving "displacement" of all molecules, as still overwhelmingly believed. They materialize by nucleation and growth as well, no matter how minute the seeming "distortion" could be in some cases.

The OR can be rigorous or not. There are two circumstances when the OR is rigorous. One is in pronounced layered crystal structures. A layered crystal consists of strongly bound, energetically advantageous two-dimensional units − molecular layers − usually appearing almost unchanged in both phases. The interlayer interaction in these crystals is relatively weak by definition. The difference in the total free energies of the two structural variants is small. (This is why layered crystals are prone to polymorphism). The phase transition is mainly resulted in changing of the mode of layer stacking, while layers of the new phase retain the previous direction. Nucleation occurs in the tiny interlayer cracks. Given the close structural similarity of the layers in the two polymorphs, *this nucleation is epitaxial* (oriented) due to the orienting effect of the substrate (opposite surface of the crack). Another case of the rigorous *epitaxial nucleation* is when the unit cell parameters of the polymorphs are extremely close (roughly less than 1%) even in non-layered crystals, as in the case of ferromagnetic phase transition of iron (Section 4.2 in [14]).

*Epitaxial* phase transitions exhibit themselves in a specific way: the OR is preserved, the x-ray Laue-patterns of the phases appear almost identical, hysteresis can be small and easily missed, so is latent heat, etc. Without a scrupulous verification, the phase transitions may be taken for "instantaneous",

"cooperative", "second-order", *etc*. In the case of the 1924 x-ray study of lead [46] it was the cause of the erroneous conclusion on the structural identity of the pre-superconducting and superconducting phases.

### 4.4. Identification of the Phase Transition Order

The Landau's definition of seconds-order phase transitions provides distinctive details to the option *'B'* in Section 2. Any deviation from its strict conditions, however small, would violate their physical meaning. In each such case the transition would be not a second order, but first, that is, materialized by a reconstruction at interfaces rather than homogeneously in the bulk. Identification of a second-order phase transition must not be "approximate"; in this respect it should be noted that its "pure" example has never been found. The core of the definition of second-order phase transition is the absence of *any* hysteresis; this alone had to prevent the normal-superconducting phase transitions to be classified as such, for all of them exhibit hysteresis, large or small.

In order to distinguish between the two kinds of transition a reliable indicator must be used to tell whether the process is localized at interfaces, where transfer of the material between the coexisting phases takes place (first order), or it is a qualitative change involving the whole bulk simultaneously (second order). The reliable indicators of *first-order phase transitions are interface, heterophase state, hysteresis of any physical property, latent heat*. Any one is sufficient, for all four are intimately connected. Any one of them will guarantee all being present. Detection of a two-phase coexistence in any proportion at any temperature would prove the transition being first order.

Thus, identification of a first-order transition is simple and definite. Not so with second-order phase transitions. Physical state of the two phases at the "critical point $T_c$" must be identical. Proving a case requires to show that all the above first-order indicators are absent. Some can be overlooked or remain beyond the instrumental capability. The same is true for "jumps": their seeming absence cannot serve as a reliable indicator as well. It is the small or undetected "jumps" that were frequent source of erroneous classification. The least reliable but frequently used indicator of second-order phase transition is absence of latent heat. Not only the latent heat can be too small for detection, it is prone to be confused with a heat capacity even when observed, as demonstrated above.

## 5. Normal – Superconducting Phase Transitions: Common Treatment

### 5.1. Assigning the "Order"

Some phase transitions, both the normal-superconducting and otherwise, are still being frequently called "second-order" in the literature. When assigning "second order" to a phase transition, it is a common practice to use only a single rational while disregarding evidence unequivocally indicating the first-order transition. It is always a blunt statement lacking sufficient argumentation. Different criteria are used. In superconductors it is the seeming crystal-structure phase identity. Or it can be latent heat as, for example, the claim [47-49] that a normal-to superconducting transition changes from first order (with a latent heat) to second order (without latent heat) with increasing magnetic field. Sometimes it is a missed (or "too small") "jump" of some property. Sometimes it is the narrow range of transition designated to be a "critical point $T_c$". Finally, if it is possible to be more wrong than all the above, it is the "specific heat lambda-anomaly", because it is not a specific heat, but the latent heat of polymorphic transition.

Proper verification of the remaining "second order" phase transitions will turn them to first order. In fact, a slow, but steady process of second-to-first-order reclassification is going on. The ferromagnetic phase transition in Fe at ~769 °C is a glowing example. For decades, it was regarded as best representative of second-order phase transitions. The two involved crystal phases, just like in superconductors, were believed to be identical and even marked as a single crystal phase. No "jumps" were ever reported. But it was shown (Sections 4.2 and 4.7 in [14]) that a change of the crystal structure in that phase transition had been overlooked. What's more, it was explained [40] that *all* ferromagnetic phase transitions are "magnetostructural". The latent heat of the ferromagnetic phase transition in Fe was ultimately recorded [50].

### 5.2. "Weakly First-Order" Phase Transitions

The "order" problem of normal – superconducting phase transitions is still unsettled in the current literature. It became even more cumbersome due to the better awareness that they do not fit second order as defined, considering that they occur over a temperature interval rather than in a fixed point and exhibit hysteresis. Phenomenologically, they matched to the first order better. Without sacrificing scientific purity they hardly could be regarded a "critical phenomenon" and, as such, suitable for treatment by statistical mechanics. Yet, the out-of-date claims that these phase transitions are second order on the ground that there is no latent heat (which is factually incorrect) are not extinct completely. A single answer to whether all normal – superconducting phase transitions are a first or second order is not existing anymore: some are now called first order, others are called second order. Moreover, as already noted, there are claims that one and the same phase transition can have different "order" under different conditions.

Trying to salvage the area of application of statistical mechanics, the theorists resorted to a compromise – "*weakly first order*" phase transitions – and treated them as *second order*. No attempt is made to look into the irreconcilable physical difference between the molecular mechanisms of the first- and second-order phase transitions. Instead, the approach is entirely phenomenological. As such the two mutually exclusive processes became reconcilable. Definitive information about the "weakly first-order phase transitions" is absent. In general, the effects accompanying phase transition – change of crystal structure, transition range, energy jumps, latent heat, discontinuities in specific volume, hysteresis – should be "small". How small? It depends on the researcher's perception. If the decision is made that a first-order phase transition is "weakly first order", then its treatment as a *continuous* is considered permissible.

Application of statistical mechanics assumes that the process under investigation is based on a fluctuation dynamics involving all particles in the bulk simultaneously. But the physical process in the first-order phase transitions is opposite, no matter how "weakly" they are. They materialize by restructuring that takes place only at the interface between the two contacting crystal phases. The process involves only a single particle at a time, while the crystal bulk on the both sides of the interface remains static (Fig. 4). Obviously, the application of statistical mechanics to the non-statistical process is quite inappropriate.

The following simple rule will eliminate all confusion regarding the order of any phase transition in the solid state. If the process is homogeneous in the bulk, then it is *second* order, and if it is localized on interfaces, it is *first* order. We can assure the readers that the latter will always be the case – even in the "weakliest" first-order phase transitions.

## 5.3. "Anomalies" at $T_c$: Distortions, Displacements, Deformations

Since 1924 the *concept of structural identity* (CSI for brevity) of normal and superconducting phases is impairing the investigation of superconductivity. It is instructive to trace it chronologically.

– (1924) The CSI, inferred from the investigation of a single superconductor (lead) [46] was unconditionally accepted by everyone as valid for all superconductors, thus unwittingly impairing the ensuing scientific research.

– (1934) Rutgers [51] published his work, applying Ehrenfest's second-order phase transitions to superconductors. The CSI plaid a major role in that endeavor. As we presently know, the phase transitions in question are not second order.

– (1952) The CSI is holding firm. Nobody still expressed doubts that it might be incorrect or, at least, not covering all superconductors. It diminished the value of Laue work on superconductivity [52], especially his thermodynamic theory of normal-superconducting phase transitions where the CSI was quite essential.

– (1972) Publication of three articles by H. R. Ott [53-55] where it was revealed that single crystals of lead, aluminum, zinc and gallium undergo dimensional and volume changes upon their normal-superconducting phase transitions. That work ruined the CSI by proving that the normal and superconducting phases are different crystal structures.

– (1987) Evidence of "structural distortion" at $T_c$ in some high-temperature superconductors [56, 57].

– (1991) The CSI is alive, however; the Ott work did not have the impact it deserved. P. F. Dahl [1] still describes CSI as decisively proven. There is no indication that anyone claims otherwise.

– (1992) In the frenzy of the experimental investigations of the "high-temperature" superconductors after their discovery in 1986, the CSI may seem to silently left behind. The question whether CSI is right or wrong is not raised in the experimental literature. Instead, experimental studies were aimed at finding what changes *in* the crystal structure occur at the transition into superconducting state. A number of those studies, performed on different superconductors, in which several highly sensitive experimental techniques were employed, are collected in the book "*Lattice Effects in High-$T_c$ Superconductors*" [58]. Ample evidence is presented that *there are* certain changes at $T_c$. These changes are reported as lattice *distortions*, thermal expansion *jumps*, atomic *displacements* from regular lattice sites, local *displacements* from average crystal structure, *changes* of some inter-atomic distances, etc.

– (1992-2015) The reports on a "structure anomaly", "lattice distortion", or alike, at $T_c$ in different high-temperature superconductors continued to appear, e. g. [59-62], suggesting that the phenomenon is general to all superconductors. Treatment in the literature of all those distortions, jumps and displacements at $T_c$ has a general trait: the superconducting crystal structure is actually regarded as the same pre-superconducting crystal structure, only somewhat modified. Even though the CSI is not openly present there, its shadow invisibly affects the scientific thinking. The following question could (but did not) arise: why is it the superconducting phase that becomes distorted? Would not be it also fair to regard the pre-superconducting phase becoming distorted in case of a superconducting-to-normal phase transition? There is no need to answer. Neither of the two phases is distorted. The phase transitions in superconductors and non-superconductors alike do not occur by displacements, distortions or deformations. They proceed in both directions by a nucleation-and-growth. Notwithstanding how minute the structural changes could be, the phase transition is a *replacement*, and not *modification*, of the old structure. The two crystal structures, no matter how similar, are built according to the rules of crystallography and minimum free energy.

(2015:) Disregarding the Ott's experimental results and the numerous reports on structural "distortions" at $T_c$, the CSI is still holding in the minds of some theorists. Recent examples: (1) "High-resolution X-ray data show no change in crystal structure at $T_c$, indicating no first-order transition" [63]. (2) "Measurements show that at the superconducting transition there are no changes in the crystal structure or the latent heat release and similar phenomena characteristic of first-order transitions". "The complete absence of changes of the crystal lattice structure, proven by X-ray measurements, suggests that ..." [64]. (3) "When a superconductor is cooled below its critical temperature, its electronic properties are altered appreciably, but no change in the crystal structure is revealed by X-ray crystallographic studies" [65].

## 6. Justy and Laue Were Correct, but Ignored and Forgotten

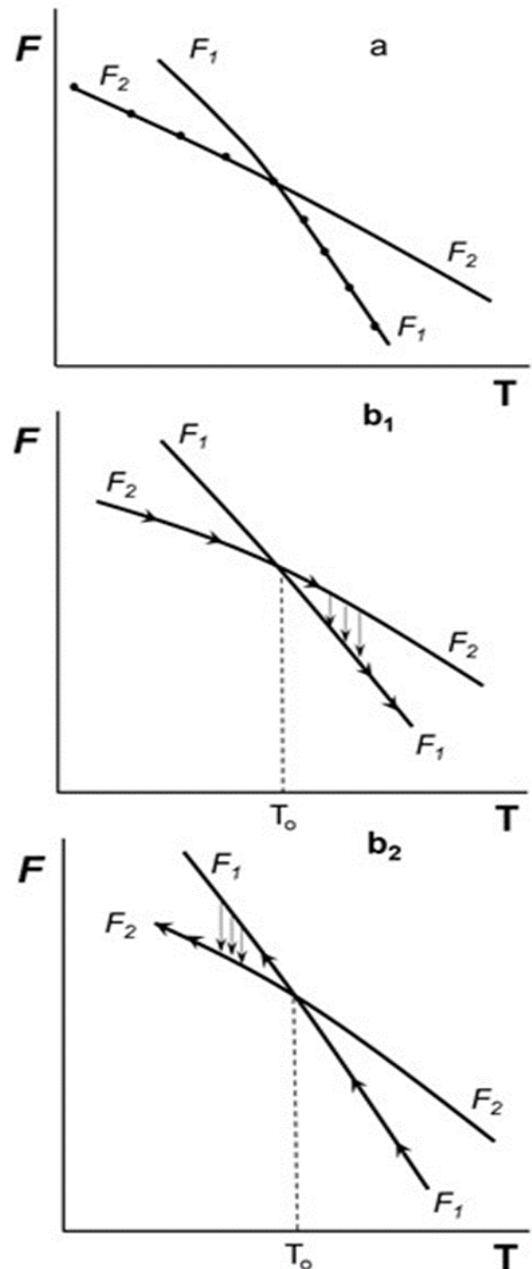

Fig. 5. a. Two curves of free energy, F1, and F2, representing two phases involved in a first-order phase transition. The dotted portion of each line indicates the thermodynamically stable phase. The scheme does not exactly reflect the actual behavior of the real system. Phase transition will not proceed without lowering the free energy of the system and, therefore, will not occur at the crossing point where F1 = F2. **b₁, b₂**. Behavior of a real system when hysteresis provides a driving force for phase transition to occur. Range of transition is also schematically taken into account. **b₁** heating, **b₂** cooling.

### 6.1. Why Second-Order Phase Transitions are Impossible

Second-order phase transitions were theoretically created by Ehrenfest [2] as follows. He used the thermodynamic free energy $F$ as a function of other thermodynamic variables. Phase transitions were classified by the lowest $F$ derivative which is discontinuous at the transition. *First-order* phase transitions reveal a discontinuity in the first $F$ (T,$p$) derivative (temperature T, pressure $p$). *Second-order* phase transitions are continuous in the first derivative, but show discontinuity in the second derivative of F (T, p): there is no latent heat, no entropy change, no volume change, and no phase coexistence.

Justi and Laue in their lecture (delivered by Laue) at the September 1934 meeting of the German Physical Society in Bad Pyrmont [66] rejected the possibility of second-order phase transition. The presentation was of a general significance, not limited by superconductors. Laue made strong thermodynamical arguments against the physical realization of Ehrenfest's criteria for second-order phase transitions. He analyzed $F$ (T) for two competing phases. In case of a first-order phase transition the two $F$ (T) curves intersect (Fig. 5). In case of a second-order transition these curves do not intersect, but osculate at the alleged critical point and then either separate again (Fig. 6**a**), or merge (Fig. 6**b**). In both cases the phase represented by the lower curve remains stable in all temperatures, unchallenged at the point of osculation. It was a proof that *second-order phase transitions cannot materialize.*

There was only a single opposition to the Justi and Laue analysis. It surfaced much later and was marked by a dishonesty. Thirty years after the meeting in Bad Pyrmont, Gorter [67] published a review with recollections of Justi and Laue presentation. He stated that he attended this meeting and pointed out to Laue that his diagrams are incorrect. He wrote: "With Laue, Keesom and I had rather tenacious discussions". This is not true. The published discussion of the presentation reveals that Gorter only suggested that order of phase transition may not necessarily be an integer number. This comment, let alone being unsound, was not significant for the essence of the Justi and Laue presentation and could hardly be called a "tenacious discussion". Then Gorter went on saying in the review that by introducing an "internal parameter indicating a degree of superconductivity" he and Casimir demonstrated in a short communication that the Justi and Laue osculating diagram is incorrect. Indeed, Gorter and Casimir have presented their short-lived hypothesis on "degree of superconductivity" by publishing the same article in two different journals and a book chapter [68-70].

However, there was no critique of Justi and Laue presentation, nor the osculating diagram. No wonder Laue never responded to the Gorter's critique, since there was no one. Gorter completed his recollection on the topic by saying "It is clear that in this model there is no place for Laue and Justi's objection". We have two notes to that statement. (1) The reverse is true: there is no place for a theoretical model that is not in compliance with thermodynamics. (2) Calling the proof by Justi and Laue an "objection" reduced the result of the rigorous thermodynamic analysis to the rank of an opinion. Laue did not change his position in the later published book and the article on superconductivity [52, 71]. The Gorter review was published four years after Laue death, so Laue could not already respond to it.

*Justi and Laue must be credited for finding that second-order phase transitions cannot exist. Unfortunately, their critically important contribution was damaged by Gorter by providing a pretext to ignore it, as did L. Landau, creator of the theory of second-order phase transitions.*

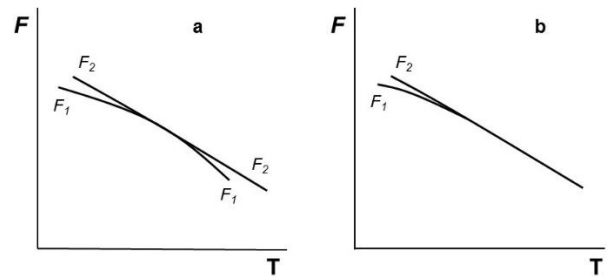

Fig. 6. Pairs of curves of free energy, $F_1$ and $F_2$, involved in the alleged second-order phase transition: **a** in the case of touching; **b** in the case of merging.

### 6.2. Laue's Further Efforts

At the time when Justy and Laue tried to persuade the contemporaries that second-order phase transitions cannot exist, the *nucleation-and-growth* mechanism of solid-state phase transitions, as described in Section 5, was not yet discovered. In spite of their efforts, a general belief has been firmly established that second-order phase transitions are real, while the normal – superconducting transitions were their primary case. In that environment Laue in his 1952 book [52] concentrated narrowly on *normal – superconducting* phase transitions. Not using the term "second order", he did his best to show that they are not.

To make a strong case a positive description of non-second-order phase transitions in superconductors had to be given. There was a major hurdle in the way: the "structural identity" of the phases. Like everyone else, Laue erroneously accepted it as a fact. We read in his book (p. 4): "The transition from normal to superconductor does not change the form or the volume of the specimen; its lattice remains the same not only in is symmetry but also in its three lattice constants. This was proved for the lead by Kamerlingh-Onnes and Keesom using X-ray analysis". It had taken another 20 years before it was shown that the form and volume of single-crystal superconductors do change.

In the absence of a consistent picture of "non-second order" phase transitions, Laue could only demonstrate that *normal – superconducting* phase transitions are incompatible with being second order. He argued: "The [normal – superconducting] transition is by no means continuous", and "In the present case one solid phase changes directly into the other, which is in contact with it", and "The 'intermediate state' almost always appears, a mechanical mixture of normal and superconducting parts", and "The transition from the superconducting to the normally conducting state requires heat: the converse process liberates heat". These properties were quite sufficient to reject its second-order nature. In retrospect, we deal here with the inalienable attributes of the nucleation-and-growth mechanism of solid-state phase transitions. But that was not enough for others: for them a normal-superconducting phase transition with its "structural identity" still looked like "second order". The misconception survived in some places to the present days. A question whether phase transitions in superconductors are the first or second order is not being settled definitely. Even worth: they incresingly seemed compatible with neither. But we eliminate this problem now: they all have the nucleation-and-growth mechanism.

In his interpretation of *normal – superconducting* phase transitions Laue was way ahead of his contemporaries, but still far from understanding the phenomenon. For example, phase transitions by "jumps" were not rejected, the word "nucleation" was not even present in the considerations, the hysteresis was incorrectly ascribed to relaxation effects, the cause of the transition range and uncertainty in the location of transition temperature remained unknown, *etc*.

### 6.3. Forty Years Later: They are not Found

Since 1934, the existence of second-order phase transitions became an unbreakable dogma. Forty years later, in spite of the unfavorable environment, claims of their non-existence were resumed (Mnyukh [14, 30, 36, 41, 44]), growing more categorical with years.

(1973) "Polymorphic transitions of second- or 'higher' order are not found"; "As for the different classifications of the phase transitions in solids, … they will, most probably, simply be reduced to polymorphic transitions of the *epitaxial* or *non-epitaxial* type".

(1978) "It is now impossible to find even one well-documented example of a second-order polymorphic transition!"

(2001) "Only one problem with the classification by first and second order will remain, namely, to find at least one well-proven crystal phase transition that would be not of the first order".

(2013) "True second-order phase transitions will never be found. The first / second-order classification is destined to be laid to rest."

(2014) "Essential result of [our] studies was the conclusion that second-order phase transitions do not exist".

These "heretical" statements resulted from a comprehensive study of solid-state phase transitions undertaken from $1960^{th}$. They led to the discovery of the *nucleation-and-growth* molecular mechanism of polymorphic phase transitions (outlined in Sections 4.1, 4.2), universal to all phase transitions called "structural" in the literature. It was followed by the discovery of its *epitaxial* variation (Section 4.3) which is usually mistaken for "second order" phase transitions. No reasonable niche to accommodate second-order phase transitions remained.

A major milestone was the establishment of a structural nature of *ferromagnetic* phase transitions, considering that they have been used in the literature as the last resort for second-order phase transitions. It was shown [40] that all ferromagnetic phase transitions must materialize by a crystal rearrangement at interfaces by the *epitaxial* mechanism. The same is true for ferroelectric phase transitions. Adding to the above are *order-disorder* phase transitions. They were demonstrated to proceed also by a nucleation and growth of the orientation-disordered crystals in the original crystal medium [32]. The "heat capacity λ-anomaltes" almost completed the picture. They used to be the reason for the introduction of the second-order type of phase transitions (see Section 3.1), but turned out to be a *latent heat* of the nucleation-and-growth.

Only the phase transitions in superconductors were not directly analyzed previously. Not any more: we have shown in this article that they fall in line with all other solid-state phase transitions. There is enough evidence that second-order phase transitions do not exist at all, being a product of excessive theoretical creativity.

## 7. Example

This section is to illustrate how familiarity with the molecular mechanism of solid-state phase transitions would eliminate serious misinterpretations and wrong conclusions detrimental to the science of superconductivity.

In the considered work [49] the specific heat of a high-quality single crystal of superconductor $ZrB_{12}$ was measured with and without an applied magnetic field. Application of magnetic field of increasing strength shifted the temperature of normal – superconducting phase transition (~ 6 K in zero field) down. The recordings of specific heat $C = f(T)$ were typical "λ-anomalies". The recorded value was a combined contribution of specific heat $C$ and latent heat $Q$, even though only $C$ was shown on the ordinate axis. Without magnetic field the λ-peak was high and thin and would traditionally be identified with second-order transitions. This time, however, it was assigned first order. As the applied magnetic field was increased, the λ-peaks became lower and wider until degrading into a diffuse hump at ~ 1 K. This hump was assigned to represent a second-order phase transition. It was concluded that *"the normal-to-superconducting state transition changes from the first order (with a latent heat) to second order (without latent heat) with increasing magnetic field"*. Nothing was said about the non-integer "order" of the intermediate cases between these two extremes. *Hysteresis* and *range of transition* were present in all recordings, but had no role in that claim. In fact, they were enough to invalidate the presence of a second-order transition, but the only used criterion for assigning the "order" was latent heat.

But the most revealing misinterpretation that nullifies the theoretical claims of that work, including about a type of the superconductor, was in the wrong identification of the *latent heat* and *the specific heat* in the experiments. It is due to this mix-up the improbable phenomenon of turning first-order phase transition into second order (and vice versa) entered the scientific literature. According to the authors, "the latent heat [is] the area below the specific-heat peaks". *But it is the λ-peaks that are a latent heat*. The latent heat peaks rest, as on a baseline, on the curve delineating specific heat over the range of phase transition (Fig. 7).

## 8. Structural Approach to Phase Transitions in Superconductors

Nonexistence of second-order phase transitions means that all normal – superconducting phase transitions are first order. Considering that being first order is equivalent to materialize by changing the crystal structure, and the change has the *nucleation-and-growth* mechanism, we have arrived at the following conclusions.

► The crystal structures of the normal and superconducting phase are not identical. The inference from the x-ray data made in the early years of superconductivity that there was no change of the crystal structure at $T_c$ was in error, probably resulted from the experimental imperfections and the *epitaxial* type of the investigated phase transition when the difference is difficult to detect.

► The phase transitions in question should be approached as being primarily *structural*. It is important to correctly identify the cause and the effect. The cause in every solid-state phase transition is crystal rearrangement, and the effect is the physical properties of the new crystal structure. In our case the new property of the resultant crystal is its superconducting state. In order that a phase transition could occur, two crystal versions of almost equal free energies $F_1 \approx F_2$ must exist. The towering component in the both $F_1$ and $F_2$ is the energy of chemical bonding. The contribution due to superconductivity can only affect the balance between $F_1$ and $F_2$ in favor of one or the other, but not to cause a phase transition if the $F_1 \approx F_2$ does not exist. Therefore, it is the structural phase transition that gives rise to superconductivity, and not *vice versa*. *Superconductivity is a property of a specific crystal structure*.

► In view of the fact that a *normal – superconducting* phase transition has the same universal *nucleation-and-growth* molecular mechanism as all other solid-state phase transitions, it bears no relation to the physical nature of superconductivity. That does not mean a *comparison* of the pre-superconducting and superconducting structures is useless. Quite opposite: it could bring valuable information, especially in the epitaxial cases. Indeed: why does such a minor crystal-structure change produce that drastic change in electrical conductivity? Because of the belief in the structural identity of the phases, this direction of research had been excluded up to the time of discovery of the high-temperature superconductivity in 1986, and is still negatively affecting scientific thought.

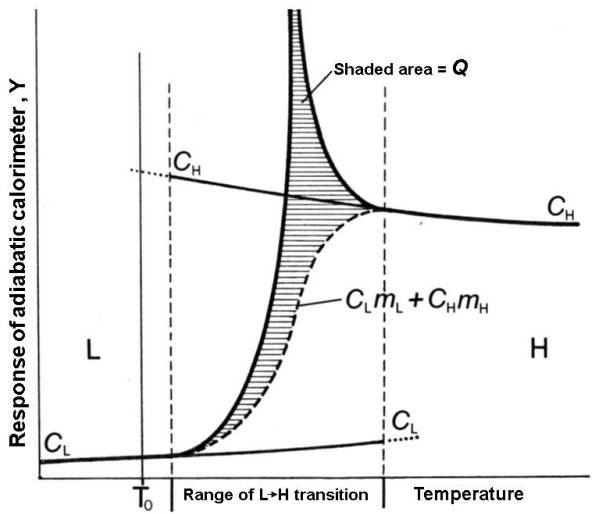

Fig. 7. Heat capacity $C$ and latent heat $Q$ in the range of transition between phases marked L and H. The drawing features: Two separate overlapping plots $C_L(T)$ and $C_H(T)$ would represent the specific heat of each phase. The S-curve (dashed bold) represents an amalgamated plot of two independent contributions $C_L$ and $C_H$ from the coexisting phases L and H, taking into account their mass fractions $m_L$ and $m_H$; $m_L + m_H = 1$. The bold-faced curve with "λ-anomaly", which would be produced by calorimetry. It is composed of both $C_L$ and $C_H$ contributions (dashed curve) and a superstructure of the latent heat $Q$ of the phase transition. (Reproduced from [14]).

► The phenomena believed to be a deviation from the "ideal" equilibrium *normal – superconducting* phase transition – range of transition, hysteresis, phase coexistence, uncertain transition temperature – receive now final explanation. They are simply the inalienable properties of the universal *nucleation-and-growth* mechanism, which is the only way solid-state phase transitions materialize.

► The problem of exact temperature of normal – superconducting phase transition (and any other phase transition for that matter) is also clarified. It is not the temperature at which the resistance is one half of the value just before the drop, as it is commonly defined. Neither it is located at the foot of that curve, as Laue suggested. From a thermodynamic point of view, it should be the position of $T_o$ when $F_1 = F_2$, but it is not directly achievable in experiments, considering that any measured temperature will be either above or below the $T_o$ due to hysteresis. The proper (but still approximate) position of the $T_o$ can be found only by extrapolation of the measured temperatures in heating and cooling runs. The $T_o$ will be somewhere in between; it is the only temperature constant characteristic of that particular phase transition. Calling the *observed* temperature "critical temperature (or point) $T_C$" is incorrect: it is not a fixed point. Crystal growth is not a "critical" phenomenon eligible for treatment by statistical mechanics.

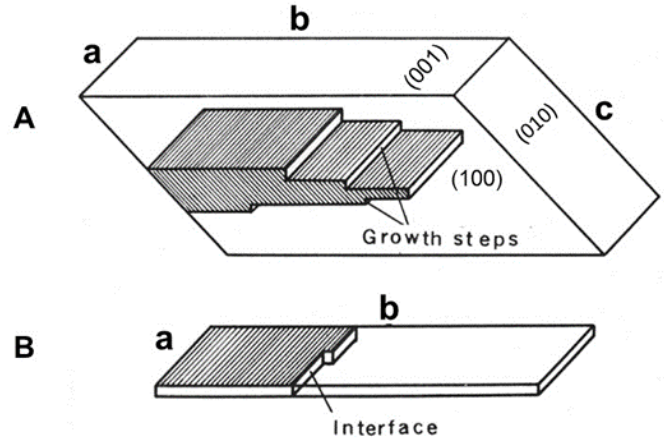

Fig. 8. **A**. Schematic illustration of phase transition in a layered crystal. It proceeds by growth of the wedge-like crystals of the alternative phase within the initial single crystal. This is an epitaxial type of phase transition – when the molecular layers in the two phases are very similar and retain their orientation in the transition. **B**. Every step of the growing new crystal phase propagates by shuttling movements of kinks along the step.

► The realization that superconducting state is a *property* of a specific crystal structure opens the experimental possibility to observe superconductivity at room temperature. All it requires is to prevent the superconductor in its $\rho = 0$ state from structural rearrangement when the temperature is rising. As indicated in Section 4.1, a polymorphic phase becomes practically stable in all temperatures if its lattice is of a sufficiently high quality. Therefore, in order to have a superconducting crystal at room temperature, the only problem would be to grow it defect-free at the temperature when it is still in the superconducting state. The practical stability of its superconducting state at the temperatures well exceeding the $T_o$ will depend on the degree of its perfection.

► The foundations of the Ginzburg-Landau theory of superconductivity are questioned. (1) It erroneously implies that the mechanism of a *normal – superconducting* phase transition is specific to that type of phase transitions. However, it is general to all solid-state reactions and, therefore, has nothing to do with the resulting superconductivity. (2) The theory in question is based on the Landau theory of seconds-order phase transitions. Not only they are nonexistent, their alleged molecular mechanism is antithetical to that of real phase transitions.

► The realization that normal → superconducting phase transitions occur by a reconstruction of the crystal structure, and that it has the *nucleation-and-growth* mechanism, opens an opportunity to correlate recording of electric conductivity with the physical process in the investigating object. Because high-T superconductors typically have a layered structure, we illustrate it with a layered single crystal. Fig. 8 shows schematically the process of crystal rearrangement during a phase transition in layered single crystals of hexamethyl benzene [36]. That it is reproduced from a work not related to superconductivity is of no significance.

The single-crystal shown in 'A' is brittle, easily cleaved into thin plates parallel to *(001)*. If it is of good quality, the new phase (now we assume it to be superconducting) is developing by epitaxial growth of a wedge-like crystal within it with its flat faces parallel to the *(001)* cleavage planes. The growth steps shown in 'A' move over one another toward *(010)* face. Their movement is by kinks (shown in 'B') shuttling between the *(100)* faces. In the absence of this picture, the researcher will not be able to explain why steepness and shape of the electrical resistivity curve $\rho = f(T)$ depend more on the position of the measuring electrodes than on the physics of the phenomenon. Indeed, if the sample is confined between the electrodes attached to the *(100)* faces, the resistivity drops to zero as soon as the very first shuttling kink makes a connecting string. The $\rho$ drop could be so steep, that its difference from $90^0$ might not be even measurable. At this point, while the $\rho = f(T)$ curve shows the normal → superconducting transition already completed, it is barely started, considering the almost whole crystal is still in non-superconducting phase. But some steep before $\rho$ becomes zero can be expected for the measuring electrodes attached to the *(010)* faces, and even more so for the electrodes attached to the *(001)* faces. In the latter case, the $\rho = 0$ is reached only when the almost whole frame of the original sample is filled by superconducting phase. This example may serve as a warning not to assign too much value to the delineation of the $\rho = f(T)$ curves.

## 9. Conclusions

1. A comprehensive evidence is collected, both theoretical and experimental, that second-order phase transitions are non-existent in nature and even unsuited to approximate the real phase transitions: (a) The "λ-anomaly" in liquid *He* is not *a heat capacity*, but *a latent heat of the phase transition*; (b) The thermodynamic proof by Justy and Laue that second-order phase transitions cannot materialize is valid and true today; (c) The *epitaxial* phase transitions are misinterpreted as the second order; (d) Not a single well-documented second-order phase transition in solids exists.
2. Since all solid-state phase transitions are first order, the first/second order classification is not needed.
3. The first-order phase transitions constitute an *overall replacement* of the crystal structure, notwithstanding how minor the change could be. They are not a distortion/deformation of the original structure, or displacements of its certain atoms. It follows that, contrary to the common belief, superconducting phase has an individual crystal structure not identical to that before the transition.
4. Transitions between normal and superconducting crystal phases materialize by the general *nucleation-and-growth* mechanism of all structural rearrangements. It is not specific to phase transitions in superconductors and, in spite of the widespread belief in the opposite, it sheds no light on physics of superconducting state.
5. A comparison of pre-superconducting and superconducting crystal structures can be highly informative, especially when the structural differences are minute. It may require, however, the detailed structure analysis of the two phases with even higher resolution than presently available.
6. The Ginzburg-Landau theory of superconductivity contains a mysterious aspect. The theory was built on the assumption that mechanism of a normal → superconducting phase transition is uniquely related to superconducting state, which is not the case, for it is the same as in all other solid-state reactions. Besides, the theory assumes the transition to be second-order, which is not the case either.

7. The previously unexplained phenomena always observed in the normal – superconducting phase transitions – range of transition, phase coexistence, hysteresis, uncertainty in the position of transition temperature – are the unalienable features of the general *nucleation-and-growth* mechanism of phase transitions.

8. The clarifications provided in the present article should have positive impact on the ongoing research of superconductivity.